\documentclass{raa_twocolumn}
\usepackage{graphicx,times}             
\usepackage{natbib}
\usepackage{CJKutf8}
\usepackage{caption}
\usepackage{hyperref}
\usepackage{longtable}
\usepackage{mathtools}
\usepackage{amssymb,amsmath}
\bibpunct{(}{)}{;}{a}{}{,}

\newcommand{\orcid}[1]{%
    \raisebox{0.7ex}{\scalebox{1}{
        \href{https://orcid.org/#1}{\includegraphics[height=1.5ex]{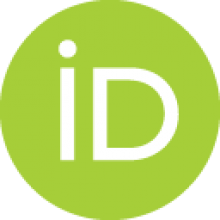}}%
    }}%
}

\begin{document}
    \title{\textit{SVOM}/VT: Overview of data processing and GRB identifications with X-band data}
   \volnopage{Vol.0 (202x) No.0, 000--000}      
   \setcounter{page}{1}          

      \author{Hua-Li Li \orcid{0000-0002-7457-4192}
\inst{1,*} \footnotetext{$*$Corresponding Authors, these authors contributed equally to this work.}
   \and Yu-Lei Qiu \orcid{0009-0007-7207-4884}
\inst{1,*} \footnotetext{$*$Corresponding Authors, these authors contributed equally to this work.}
   \and Li-Ping Xin \orcid{0000-0002-9422-3437}
      \inst{1,*} \footnotetext{$*$Corresponding Authors, these authors contributed equally to this work.}
      \and Chao Wu
      \inst{1,3}
     \and Zhu-Heng Yao
      \inst{1} 
      \and Yi-Nuo Ma
      \inst{1,4,5}     
      \and Yang Xu
      \inst{1}
      \and Pin-Pin Zhang
      \inst{1}
      \and Xu-Hui Han
      \inst{1}
      \and Jing Wang
      \inst{1}
      \and Hong-Bo Cai
      \inst{1}
      \and Da-Wei Xu
      \inst{1,3}
       \and Jesse T. Palmerio
      \inst{2}
      \and Mao-Hai Huang
      \inst{1}
      \and Jia-Li Zhu
      \inst{1}
      \and Mo Zhang
      \inst{1}
      \and Jin-Song Deng
      \inst{1,3}
       \and Bertrand Cordier
       \inst{2}
      \and Jian-Yan Wei
      \inst{1,3}
   }

   \institute{National Astronomical Observatories, Chinese Academy of Sciences,
             Beijing 100101, China; {\it lhl@nao.cas.cn, qiuyl@nao.cas.cn, xlp@nao.cas.cn}\\
        \and
             CEA/Paris-Saclay, Irfu/Département d'Astrophysique, 91191 Gif-sur-Yvette, France;\\
        \and
        School of Astronomy and Space Science, University of Chinese Academy of Sciences, Beijing 101408, China; \\
         \and
	 Institute for Frontier in Astronomy and Astrophysics, Beijing Normal University, Beijing 102206, People's Republic of China\\
        \and
	  School of Physics and Astronomy, Beijing Normal University, Beijing 100875, People's Republic of China\\
\vs\no
   {\small Received 202x month day; accepted 202x month day}}

    \abstract{
VT (the Visible Telescope) is an optical telescope onboard the \textit{SVOM} (Space-based Multi-band Astronomical Variable Objects Monitor) mission, specifically designed to detect optical counterparts of gamma-ray bursts (GRBs), study their afterglows, and select high-redshift candidates. It performs rapid follow-up observations simultaneously in two channels either via autonomous platform slewing or Target of Opportunity (ToO) observations. The science images acquired by VT and transmitted via the X-band downlink system are designated as VT X-band data. This paper provides an overview of GRB optical afterglow identifications with VT and describes the ground-based processing pipeline for VT X-band data, including preprocessing, astrometric calibration, and photometry. Up to 2025 December 3, VT has followed up 111 GRBs triggered by  \textit{SVOM} or external missions. The overall detection rate of optical counterparts is approximately 75\%. Specifically, for bursts detected by  \textit{SVOM}/ECLAIRs, the detection rate is 77\% when observed by VT within 30 minutes after the burst. A slightly higher detection rate of 81\% is achieved for GRBs triggered by external missions through rapid ToO observations with a mid-time of less than 3 hours.
}
   \authorrunning{HLL,YLQ,LPX}            
   \titlerunning{the \textit{SVOM} special issue}  
   \maketitle
%
%
%
\section{Introduction}
\label{sect:intro}

The Chinese-French collaborative satellite \textit{SVOM} (Space-based Multi-band Astronomical Variable Objects Monitor, \citet{Wei2016,Cordier2026a}) was launched on June 22, 2024. Its primary scientific objective is to detect and study various types of gamma-ray bursts (GRBs). The satellite carries four payloads: ECLAIRs \citep{Godet2026a} and GRM \citep{Sun2026} are responsible for burst triggering and alert generation, while MXT \citep{Gotz2026} and VT \citep{qiu2026} perform X-ray and optical follow-up observations.
The main scientific goal of VT is to identify and confirm the optical counterparts of gamma-ray bursts (GRBs) through rapid follow-up observations, and to provide localizations with sub-arcsecond accuracy. This serves as a bridge to enable large ground-based telescopes to perform deep photometric or spectroscopic observations, leading to redshift determination and subsequent studies of the properties of GRB host galaxies.

When ECLAIRs detects a gamma-ray burst and generates a high signal-to-noise alert \citep{Schanne2026}, it can trigger the satellite platform to slew rapidly,  and guide MXT and VT to conduct immediate follow-up observations \citep{Chen2026}. If the signal-to-noise ratio is insufficient or if an external alert is triggered, the \textit{SVOM} system can also schedule ToO observations uplinked via the Beidou system  or the S-band \citep{Bai2026,Han2026} to search for the optical counterpart or to  construct the long-term light curve. 

Over the past few decades, particularly during the successful 20-year operation of the \textit{Swift} satellite\citep{Gehrels2004}, research on the multi-band afterglows of gamma-ray bursts has significantly advanced our understanding of these extreme astrophysical phenomena, e.g.,\citep{Zhang2007,Liang2013,Dai2017,Burns2023,Saccardi2023,Gao2025}.
In particular, a detection rate of about 96\% for the X-ray afterglow \citep{2007SPIE.6686E..07B} has enabled the establishment of a rich sample of canonical X-ray light curves (e.g., \cite{2006ApJ...642..389N}). However, in the optical band, the overall detection rate remains roughly 50\%, which limited by factors such as geographic location, time zone coverage, and weather conditions, wavelength coverage and detection capability. For instance, \textit{Swift}/UVOT achieves a detection rate of approximately 48\% \citep{2023Univ....9..113O}.

Even with the relatively low detection rate,
all these efforts have  yielded a large sample of optical light curves for GRBs \citep{Kann2011}, as well as associated supernovae \citet{2025ApJ...993...20D} and kilonovae (e.g., \citet{2017ApJ...848L..27T}; \citet{2023NatAs...7..976L}). Supernova associations support theoretical models that some gamma-ray bursts originate from the collapse of massive stars \citep{Galama1998,Hjorth2003,Stanek2003}.  Such bursts can theoretically be detected from the nearby cosmos to the very early universe at redshifts exceeding 20 \citep{2000ApJ...536....1L}.

Benefiting from the featureless spectral properties of gamma-rays, high-redshift gamma-ray bursts act as beacons traversing the universe, providing a unique probe for studies of the physical conditions of the very early universe \citep{Cucchiara2011,Wang2013}. However, when the redshift exceeds 7, identification must rely on the near-infrared band,  while very early and deep upper limits in the optical could be a good indicator for selecting high redshift candidates \citep{Cordier2025}. This is one of the scientific potential contributions from VT \citep{qiu2026}. VT operates within an observational wavelength range extending to 1 micron. It can initiate observations within 5 minutes after satellite slewing, achieving a detection depth of 23 magnitudes in the early observational phase after the bursts. This significantly increases the possibility of high-redshift candidates once nondetection is determined, although optically faint  or heavily extinguished GRBs cannot be fully excluded.

On the other hand, scientific images obtained from space-based optical observations are highly susceptible to interference from cosmic rays, hot pixels, and other effects .
In particular, when the localization accuracy of the high-energy trigger is at the arcminute level, many optical candidates will fall within the error region, further complicating to identify the true counterpart. To address this, a robust strategy is to employ a combination of methods, including referencing the properties of X-ray counterparts, cross-matching with deep catalogs, combining data from two channels simultaneously, removing cosmic rays via image stacking, filtering using hot pixel lists, checking asteroid databases to exclude moving objects, and analyzing the light-curve behavior and color features of candidates. Through this multi-dimensional approach, the identification of reliable optical counterparts  to gamma-ray bursts becomes feasible.

Two types of VT data are used to identify the optical counterparts. One comprises the VT products generated onboard \citep{Cai2026}, downlinked via the VHF network with low latency and processed on the ground \citep{Cordier2026b, Wu2026a}. This approach enables the identification of optical counterparts for approximately 60\% of bursts. 
The other consists the science images transmitted via the X-band downlink system \citep{Liu2026}, which are designated as VT X-band data.

This paper presents astrometric and photometric processing of VT X-band data, focusing on the identification of GRB counterparts using VT X-band data. Since the launch of \textit{SVOM}, VT has conducted follow-up observations for 111 gamma-ray bursts up to 2025 December 3, achieving an overall detection rate of 75\% with X-band data, and has identified several high-redshift GRB candidates. One of these has been confirmed through ground-based near-infrared photometry and spectroscopy, with a measured redshift of 7.3(GRB 250314A, \citet{Cordier2025,Levan2025}).

This paper is organized  as follows. Section 2 introduces the observations of VT for the GRB program. Section 3 describes the pre-processing, astrometric calibration and photometry for the GRB afterglow. Section 4 presents the strategy  for identification of afterglow with VT X-band data. The performance of the detection and identification of GRB optical counterparts with VT X-band is discussed in Section 5.  Summary is given in Section 6. 

\section{VT Observation and data link}

\subsection{VT observations}

VT is an optical telescope with a diameter of 44 cm and a field of view of 26$\times$26 arcminute$^2$. The pixel scale is 0.76 arcseconds.  It is equipped with two 2K$\times$2K e2v frame-transfer CCDs in the VT\_B  and VT\_R channels. The two channels work simultaneously, covering the wavelength ranges of 400-650 nm and 650-1000 nm respectively. Detailed information on VT could be found in \citet{qiu2026} and \citet{Yao2026}. 
Given the anti-solar pointing constraints \citep{Cordier2026a} and Earthshine-induced stray-light limitations, high-quality VT observations are only feasible during Earth eclipse, with each orbital pass providing a ~30–40-minute observing window.

\subsection{Observations for \textit{SVOM}/ECLAIRs bursts with automatic slew}

When a GRB is detected by ECLAIRs and meets the automatic slew threshold, 
both MXT and VT automatically initiate follow-up observation within 5 minutes of the burst and continue for approximately 8 hours, covering about 5 orbits. 
During the slew and prior to platform stabilization, VT operates in chance mode with a short exposure time of 15 seconds. Subsequently, exposure times of 50 seconds are adopted to balance readout rate, noise, detection capability and the optical brightness of GRB afterglow in the early phase. 
In the initial 2 orbits, VT acquires full-frame images.
For subsequent orbits, an exposure time for each frame is increased to 100 seconds. The image size might be adjusted to window-frame centered on the MXT position, large enough to cover the MXT error region, once MXT detects the X-ray counterpart onboard. Otherwise, full-frame images are acquired with some drops \citep{qiu2026}, owing to data budget constraints. 
Throughout the observations, high-stability mode is employed to enable rapid stabilization after slewing and to enhance VT detection performance. The Finding Guider System (FGS) operates continuously, enabling the platform to maintain a stability of 1$\sigma \sim$0.85 arcseconds over 100-second intervals \citep{Li2026}.

\subsection{Observations for bursts with ToO}

 Target of Opportunity (ToO) observations are executed via telecommand through either the S-band antenna or Beidou short-message uplink channels. They are performed if no automatic slew is triggered by \textit{SVOM}/ECLAIRs owing to a low signal-to-noise (S/N) ratio, or if the bursts are detected by other satellites (e.g., \textit{Swift} or Einstein Probe(\textit{EP})), or during follow-up revisits of bursts detected by \textit{SVOM}.

The high-stability mode is optional, while the normal-stability mode with  the FGS inactive is recommended for most cases. In this mode, the platform pointing exhibits a drift of approximately 10 arcseconds per 30 minutes. The gradual pointing drift between consecutive images can help eliminate effectively thermal hot pixels and cosmic rays. An exposure time of 50-70 seconds is also recommended by taking into account of the read noise, background levels, and the impact of platform drift on VT image quality.


\subsection{X-band data link and format}

All science images are downlinked via X-band passes, with typically eight passes per day. 
All data are initially transferred from each pass to the \textit{SVOM} Mission Center, located at the National Space Science Center in China (NSSC, \citet{Liu2026}). The data are decoded into L0 data by verifying the continuous quality and other parameters, and then promptly  transmitted to the \textit{SVOM} Chinese Science Center located at National Astronomical Observatories, Chinese Academy of Sciences (NAOC), where L1 and L2 products are generated. 

For the L1 level, all science data are images in FITS format. The central 2048$\times$2048 area is the exposure region, while the left and right sides each contain 50 blank columns and 50 overscan reference pixel columns, neither of which are exposed. Similarly, above and below the exposure area, there are 20 dark current reference pixel rows each. The final product of L1 level data is defined as L1X. For this type of data, all header information is fully decoded and translated to facilitate better understanding for users. Additionally, the L1X image header includes information that characterizes image quality, such as energy concentration and background brightness. These parameters are useful for image selection, as discussed in Section 3.3.1..

The images after pre-processing and astrometric calibration are classified as L2 images (see  Section 3 for details of the pre-processing). For L2 images, only the effective exposure region (e.g., 2048$\times$2048 pixels) is retained. All the reference pixels are  removed after instrument calibrations. VT also supports the configurations for  windowed observations. It means that only the sub-window centered the images would be obtained without any reference pixels presented above. The minimum windows size is 300$\times$300 pixels.

\section{Pre-precessing, calibration and photometry}

\begin{figure}
    \centering
    \includegraphics[width=1.0\linewidth]{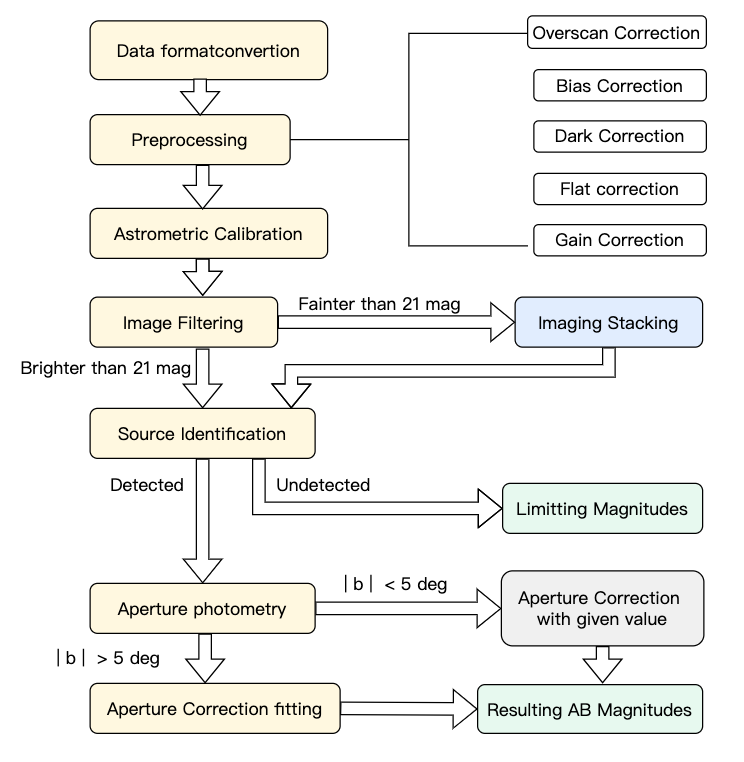}
    \caption{Flow-chart of VT X-band data processing for the GRB afterglow.}
    \label{fig:flowchart}
\end{figure}

Before identifying the optical counterpart of GRBs, data processing is required, including pre-processing, astrometric calibration, and photometric measurements. The main process is shown in Figure \ref{fig:flowchart}. The methods for pre-processing and photometric calibration are identical to those applied to non-GRB scientific targets in VT images. During the identification process, it is sometimes necessary to analyze the light curve behavior of candidates. This section provides a description on these procedures.

\subsection{Pre-precessing}

All L2 images are processed with instrumental effect correction from L1X images  in a standard manner, including overscan correction, bias correction, dark correction, and LED flat-field correction. If the readout is performed in both channels, 
a gain-matching correction for the background is applied, based on the assumption of  a uniform  background around the interface between the  two channels during readout. 

\subsubsection{Overscan correction}
\begin{figure}
    \centering
    \includegraphics[width=0.48\linewidth]{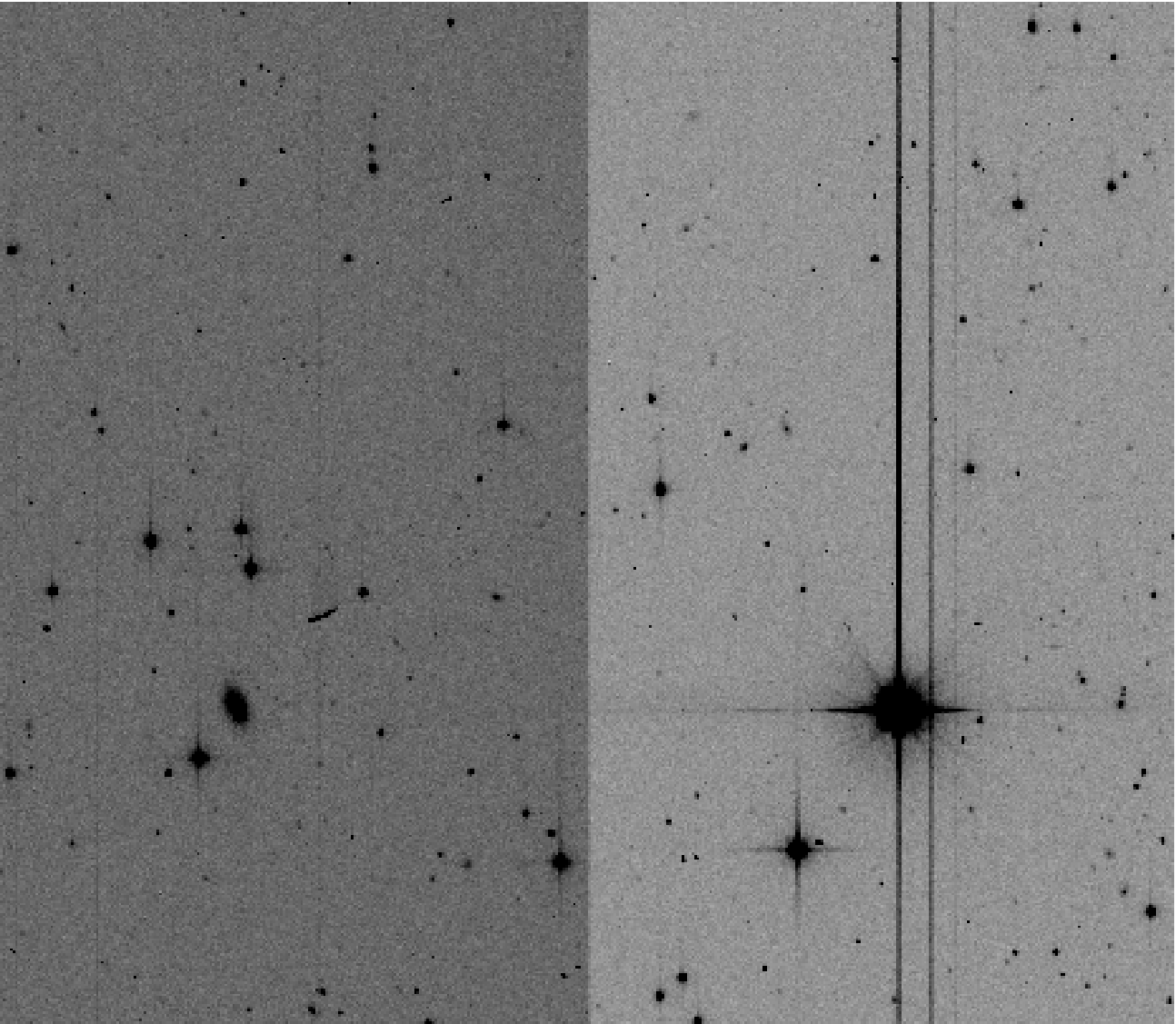}
    \includegraphics[width=0.48\linewidth]{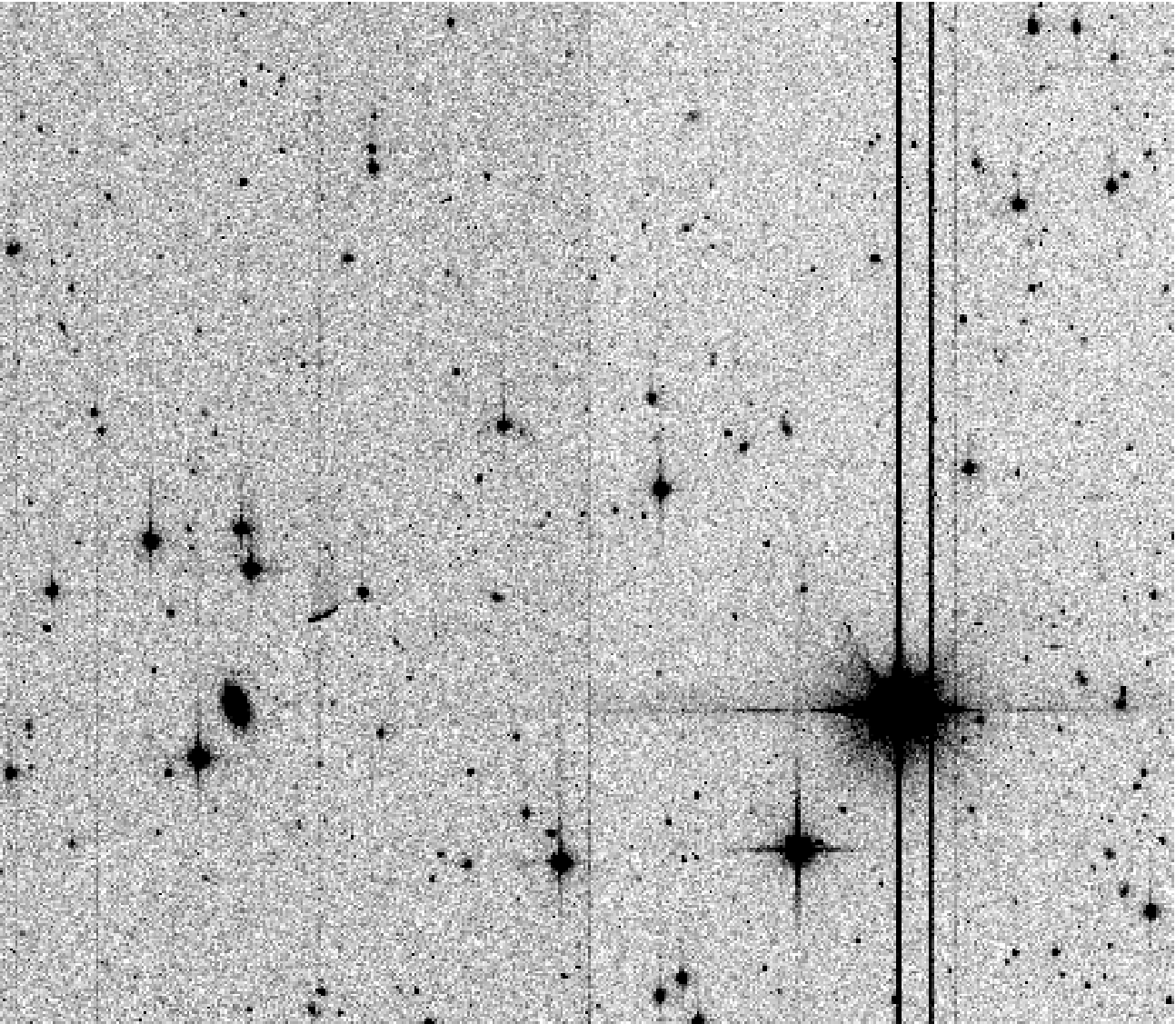}
    \caption{The comparison for the overscan correction. Left: the original image; Right:the same image after the overscan correction. }
    \label{fig:overcorrection}
\end{figure}

For overscan correction, temporal drift in the bias level can cause the image background to be either overestimated or underestimated. The resulting uneven background across left and right channels would generate spurious sources near the interface region and lead to missing stars or inaccurate photometry when the background is overestimated. Figure \ref{fig:overcorrection} shows the images before (left) and after (right) overscan correction. This correction is applied to all the images prior to other corrections. 

\subsubsection{Bias correction}

VT acquires in-orbit bias images with zero exposure time, after the  shutter in front of the two detectors is closed. 
All observations are scheduled to occur while the platform is in Earth’s shadow.
During the processing, the bias images are first examined based on their background levels, and any images contaminated by stray light from Earth are rejected. 
After the filtering all individual images  were performed for the overscan correction.
The master bias image is generated using the IRAF zerocombine package. This step is also to filter out the effects from the cosmic rays in each image. 
The bias corrections for each science image is performed with IRAF ccdpro package.

\subsubsection{Dark correction}
Similarly to the bias images, VT also acquires in-orbit dark images with  the shutter closed. The exposure time is set to 100 seconds, which is similar to the typical exposure used for science observations. During the processing, all dark images are filtered according to their background levels, and subsequently corrected for overscan effects. The master dark image is generated using the IRAF darkcombine package. During preprocessing for the science images, the master dark image is normalized to the exposure time of the corresponding science image and the IRAF.ccdpro package is then used to perform dark correction.

\subsubsection{LED flat-field correction}
A total of 18 LEDs covering six colors are  placed in front of the two detectors.
Their central wavelengths are 470 nm, 527 nm, 640 nm, 670 nm, 740 nm, and 870 nm. The blue LEDs (470 nm, 527 nm, 640 nm) are used for the VT\_B channel, while the  remaining LEDs are used for the VT\_R channel \citep{qiu2026}.
All flat-field images are pre-filtered using the same procedure applied to bias and dark-field images.
Flat-field images are then grouped by color and  combined into a master flat-field image.
Gain correction ensures uniform background from left to right across the interface as described in Section \ref{gain-correction}.
Owing to the short distance between LEDs and detectors, the backgrounds of the flat-field images exhibit non-uniform distribution. The master flat-field image is normalized using background values derived from large-scale fitting.
The normalized master LED flat-field image is used to correct pixel-to-pixel non-uniformity in the science images.

\subsubsection{Gain correction}
\label{gain-correction}

For images obtained in readout mode of "both channels" \citep{qiu2026}, the gain values of the two channels differ slightly due to electronic effects. This leads to a background difference of up to 4\% for the left and right part of images. In order to eliminate or reduce this effect, gain matching is applied for the images. The algorithm calculates the ratio of the background values for the central left and central right region.

\subsection{Astrometric calibration}

Astrometric calibration aims to determine the transformation between detector coordinates (X, Y) and the celestial coordinates (R.A. and Decl.). A custom pipeline based on the IRAF ccmap package was developed and applied to VT images, matching the brightest stars extracted from each image with the GAIA DR2 catalogue \citep{Gaia2018} as reference. 
Detailed information is provided in \citet{Yao2026}. 
In addition ASTROMETRY.NET (\citep{Lang2010}) is also used if our custom-developed pipeline fails particularly for those images with high stellar density. 
All WCS solutions are then written into the image  headers. 
The resulting astrometric accuracy is 0.01 arcsec for bright sources and 0.5 arcsec for faint sources or sources in crowded fields \citep{Yao2026}.

\subsection{Photometry}

\subsubsection{Image selection}

The initial step in performing photometry is to select images of acceptable quality. For VT, images of good quality are defined as those in which the background level and the point-source enclosed energy concentration meet the expected criteria, and the observing configuration matches the requested setup.

Typically, the background should be less than one electron per second per pixel. This holds for most observations at high Galactic latitudes. Note that this threshold may vary slightly depending on the satellite's pointing direction and orbital position. For instance, the threshold must be increased for observations where the target is close to the Moon or Earth, or lies within the Galactic plane.

The concentration of enclosed energy for point sources has remained highly stable in orbit over the long term since launch. However, image quality can be degraded by factors such as short-term platform instability or stray light from the Earth or the Moon. This quality check can be performed by inspecting the STAB\_CNT keyword in the headers of VT L2 FITS images, which records the counts per second measured by the PDPU onboard \textit{SVOM} during exposure. An image of good quality should have a STAB\_CNT value of at least 80\% of the exposure duration. 

It is important to note that images with zero-second exposure time, which are generated during orbital gaps when VT switches configurations,  are considered invalid observations and should be filtered out  at this step.

\subsubsection{Aperture photometry}

Gamma-ray bursts triggered by \textit{SVOM}/ECLAIRs are mostly located at high Galactic latitudes. The stellar density is sufficiently low that most sources are unaffected by crosstalk. Therefore, aperture photometry is applied once the targets is identified in single or stacked VT images. 

\begin{figure}
    \centering
    \includegraphics[width=1.0\linewidth]{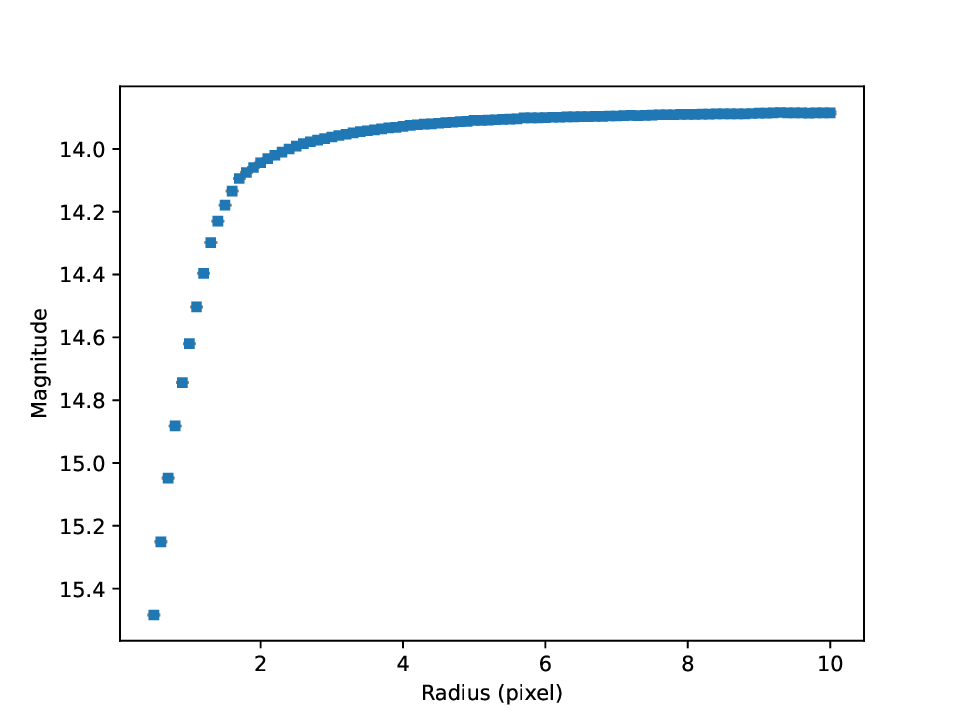}
    \caption{Growing curve in magnitude with radius for a bright source in a VT\_R band image.}
    \label{fig:growcurve}
\end{figure}

Photometry is performed in a standard manner. The first step is the determination of the photometric radius. Figure \ref{fig:growcurve} shows the growth curve of a bright star with the photometric radius and corresponding magnitudes in a VT\_R image. One can see that the magnitudes peaks at approximately 2-3 pixels from the source center. Therefore, in most cases, a radius of r=3 pixels is adopted for bright sources, and 2 pixels for faint sources. 


Aperture correction is applied,  since the aperture radius used for photometry is insufficient to encompass the total flux of bright sources.
The aperture correction can be determined using more than ten bright but unsaturated sources within the same image. As shown by the growth curve in 
Figure \ref{fig:growcurve}, a radius of r=10 pixels captures over 99\% of the flux, which is adopted as the full aperture radius for this photometric measurement.

\begin{figure}
    \centering
    \includegraphics[width=1.0\linewidth]{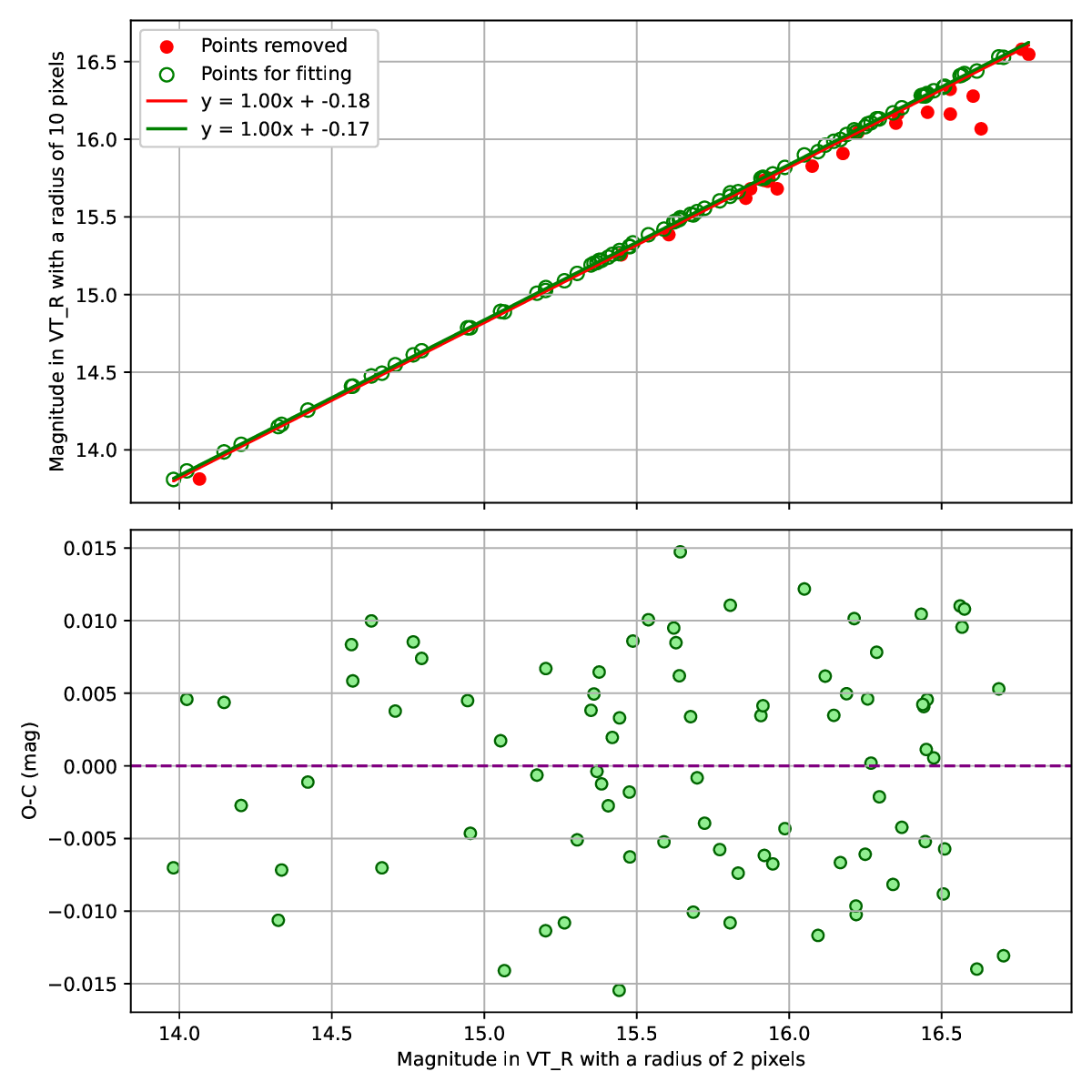}
    \caption{Aperture corrections (upper) and the precision of aperture corrections after iteration (bottom).}
    \label{fig:aperture-fitting}
\end{figure}

Figure \ref{fig:aperture-fitting} illustrates the performance to determine the value of the aperture correction through small-aperture photometry and the full-aperture radius. The fitting effectively eliminates contamination from galaxies, crosstalk from bright sources, and other unexpected effects. As shown in the lower panel of Figure \ref{fig:aperture-fitting}, the precision of the aperture correction value reaches to the order of 0.6\%.

We define $\mathrm{mag}_{\mathrm{VT}}$ as the instrument magnitude, given by :
\begin{equation}
\mathrm{mag}_{\mathrm{VT}} = -2.5*log(\mathrm{flux}*\mathrm{gain}/\mathrm{exptime})
\label{eq:flux_to_mag}
\end{equation}

The resulting magnitude after the aperture correction is then converted to AB magnitude \citep{Oke1974}  with the Equation \ref{eq:mag_ab}. 

\begin{equation}
\mathrm{mag}_{\mathrm{AB}} = \mathrm{mag}_{\mathrm{VT}} + \mathrm{Zero}_{\mathrm{VT}} + \mathrm{Cor}
\label{eq:mag_ab}
\end{equation}

$\mathrm{Zero}_{\mathrm{VT}}$ and Cor are zero magnitudes for VT in both channels with a definition of magnitude per electron \citep{Yao2026} and magnitude correction for given aperture radius, respectively. All measurements given in this work are in AB magnitude.

\subsubsection{Photometry for saturated sources}

\begin{figure}
    \centering
    \includegraphics[width=0.9\linewidth]{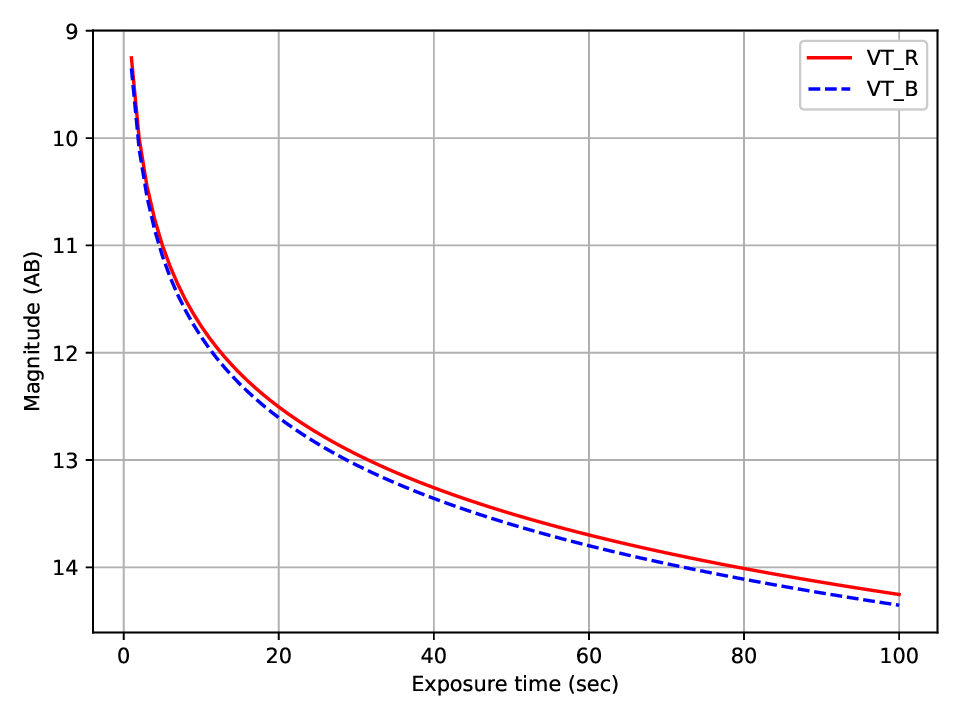}
    \caption{Saturated magnitude curves for \textit{SVOM}/VT with different exposure times.}
    \label{fig:saturated-mag}
\end{figure}

The exposure time for GRBs triggered by \textit{SVOM}/ECLAIRs is set to 50 seconds per frame during the initial 2 orbits and 100 sec for the subsequent 3 orbits. For Target of Opportunity (ToO) observations, the typical exposure time is 70 seconds for ECLAIRs off-line triggers and bursts triggered by external missions if the brightness in optical is unpredictable.  However, if the optical counterpart is too bright, it will lead to saturation in the acquired observations. 
Figure \ref{fig:saturated-mag} presents the calculated saturated magnitude curves with different exposure times for VT. Notably, with an exposure time of 3 seconds, 
the saturated magnitudes are calculated to be 10.4 mag in VT\_R and in 10.5 mag in VT\_B.  When the exposure time is increased to 50 seconds, the saturated magnitudes rise to approximately 13.5 mag in VT\_R and 13.6 mag in VT\_B.

\begin{figure}
    \centering
    \includegraphics[width=1.0\linewidth]{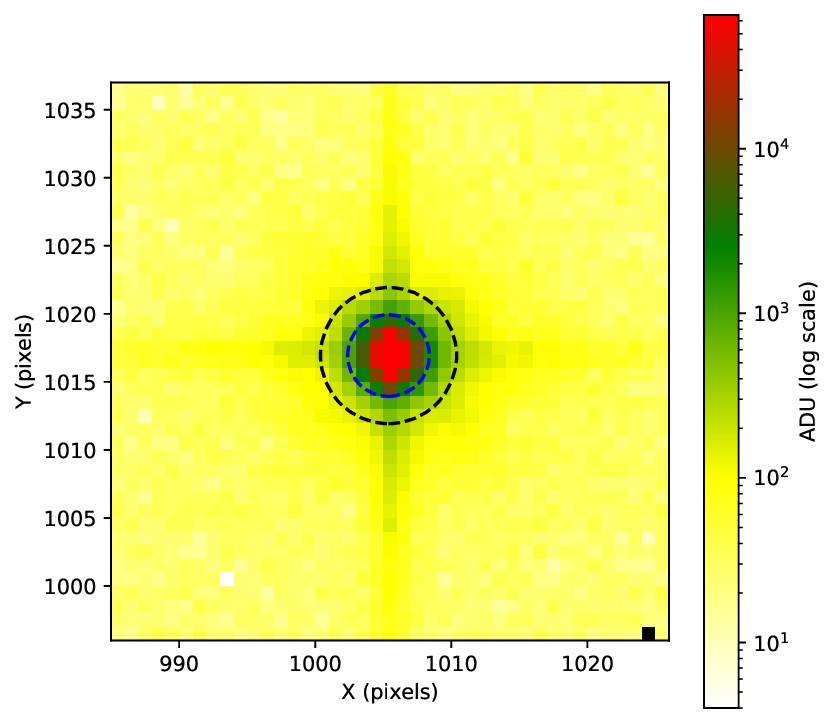}
    \caption{Illustration of radius selection for photometry of a saturated star in a VT image.}
    \label{fig:heatmap}
\end{figure}

For the photometry, we assume that the tail part of the contour profiles for the saturated star and the unsaturated bright stars in the same field follow the same trend, except for the central saturated regions. This assumption is reasonable for the uniformity of Point Spread Function (PSF) in VT images.
For the saturated star, the flux in the annulus between the inner and outer radii is measured, as illustrated in Figure \ref{fig:heatmap}.
Unsaturated bright stars in the same field of view are measured in the same way.
The calculation of the magnitude for a saturated target is then given by the Equation \ref{eq:saturated}, 


{\small
\begin{align} 
Mag_{S} = Mag_{C} -2.5 \times log\frac{Flux_{S}} {Flux_{C}}
\label{eq:saturated}
\end{align}
}%

where $Mag_{C}$ denotes to the magnitude of the unsaturated bright star measured in VT images, and $Mag_{S}$ denotes the magnitude derived for the saturated star. $Flux_{C}$ and $Flux_{S}$ refer to the fluxes in the annuli for the bright star and the saturated star respectively. 
This method has been successfully applied to the measure the optical flares from M dwarf star V* UV Cet (sb25083102, Xie et al., in prep) and 1RXS J020013.6-084106 (sb25091517, Li et al., in prep) detected by \textit{SVOM}/VT \citep{gcn41654,gcn41836}, which were both initially triggered by \textit{SVOM}/ECLAIRs \citep{gcn41614,gcn41834}.

\subsubsection{Build the light curve}

The optical light curves of gamma-ray bursts typically exhibit a decaying trend, often described by a single or broken power-law model. This decay is frequently superimposed with early-time rises or flares, which are generally attributed to the onset of the afterglow emission,  passage of the peak radiation frequency through observed band, re-activity of the central engine , or other underlying physical processes.
The distribution of the brightness can span more than several orders of magnitude at the same time post the burst \citep{Kann2011}. 

For bright cases, the photometry is performed in individual frames in order to achieve high temporal resolution, enabling the detection of afterglow onset signatures and flare activity. However, it should be noted that some measurements exhibit a sharp flux increase in only one band, which may be  due to cosmic ray hits within the aperture region. Such cases require more refined analysis.
Conversely, simultaneous flux drops observed in both channels within VT images should also be carefully examined. This may indicate poor image quality not adequately flagged during the initial quality checks, and such frames should be verified or excluded in subsequent analysis.

For faint cases, image stacking is performed. The stacking strategy primarily follows the $\Delta$T/T$\sim$1 approach, where $\Delta$T denotes the effective exposure time for the image stacking, and T represents the delay time post the burst. Given the constraints of in-orbit observations, each effective observing window lasts approximately 30–40 minutes per orbit. For images obtained during the early phase of GRB afterglow observations, the stacked time span can be as short as one orbit or even less, depending on the source brightness and specific scientific objectives.  In the later afterglow phases, stacking data from 2–3 consecutive orbits may be adopted to improve observational depth.

\begin{figure}
    \centering
    \includegraphics[width=1.0\linewidth]{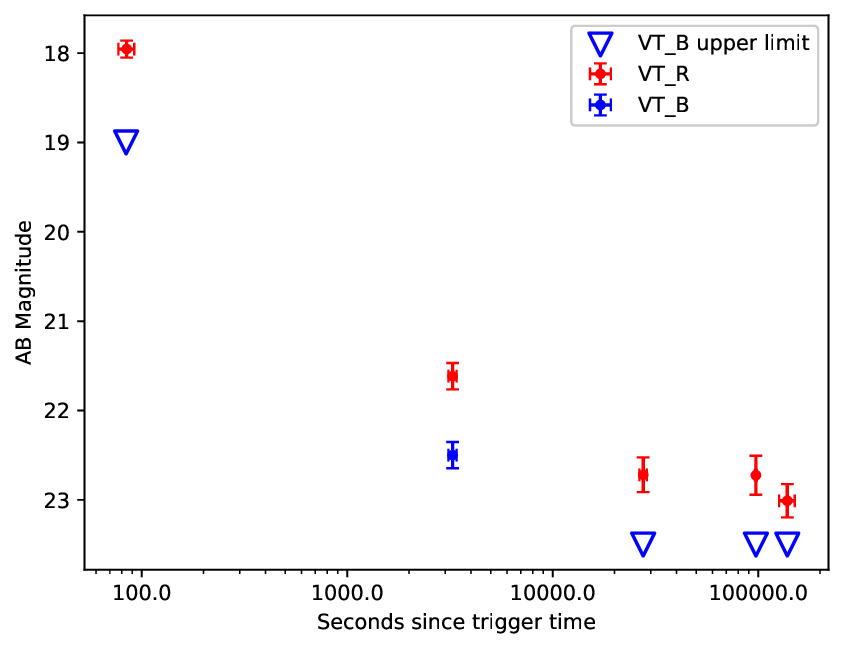}  
    \caption{Optical afterglow light curve of GRB 241209B obtained by \textit{SVOM}/VT.}
    \label{fig:lightcurve}
\end{figure}

Figure \ref{fig:lightcurve} presents the optical light curve of GRB 241209B derived from VT observations. This burst was triggered by \textit{SVOM}/ECLAIRs \citep{Xie2024}. Thanks to the slew-onboard strategy, VT initiated observations at 76.5 seconds post trigger \citep{Qiu2024}.  
The first data point in Figure \ref{fig:lightcurve} was measured from a single frame with 15-second exposure time in VT\_R at 84 seconds post trigger, while subsequent data points were generated via image stacking.

\section{Identification of optical GRB counterparts}

In the VT X-band data, the identification of GRB optical counterparts rely on three factors: the observational data sequence, the time delay relative to the trigger time, and the rapid fading behavior of the transients. The following discussion addresses three specific observational scenarios: \textit{SVOM} triggers with platform slew automatically, Target of Opportunity follow-ups observations, and the detection of high-redshift GRBs.

\subsection{Identification for \textit{SVOM}/ECLAIRs bursts with automatic slew}

VT performs observations from about 5 minutes after the bursts and lasting for 5 orbits (about 8 hours) if there are no interruptions from higher-priority observations. Considering the long-duration observations from very early time, there is a high chance to detect the reverse-shock emission or the onset of afterglows, or even prompt optical emissions in VT images. 
The subsequent behavior may show the diversity with the normal decay phase,a single power-law decay with an index of $\alpha\sim$1.0 ($f \sim t^{-\alpha}$), a jet break with more rapid fading ($\alpha\sim$2.0) \citep{2013NewAR..57..141G} or some flares \citep{2017ApJ...844...79Y} due to the long lasting central energy activities, the fluctuation of the external medium or other mechanisms.
The optical counterparts of GRB shall have the following basic characteristics:
\begin{itemize}
      \item It is an uncatalogued source compared to existing catalogs with comparable depth, except for the host galaxy, which may overlap with the afterglow position. For VT, the DESI catalog \citep{2019AJ....157..168D} is ideal as a reference in most cases.  If an uncatalogued source is detected in both VT channels, the detection is highly reliable. However,there is a small chance ($\sim$ 5\%) that the source lies in a region affected by blooming from a very bright nearby object in VT\_R, which can hinder identification or measurement in the red channel. In such cases, the counterpart may be detectable only in VT\_B.
       On the other hand, for some GRBs the counterpart may be detected only in VT\_R or undetectable in both bands, likely due to optical faintness, line-of-sight extinction \citep{Levan2006}, or high-redshift (blue-dropout) effects \citep{2010AIPC.1279..144G}. 
      
    \item The position is consistent with the trigger and the X-ray afterglow. The counterpart should be coincident within the errorbox of the gamma-ray emission and the X-ray afterglow. For GRBs detected by \textit{SVOM}/ECLAIRs, the positional uncertainty is from 2 to 10 arcminutes. The positional error of the corresponding X-ray counterpart is typically 0.5-3 arcminutes for MXT \citep{Gotz2026}, 10-20 arcseconds for  \textit{EP}/FXT\citep{2025RDTM....9..198C}, and 3-10 arcseconds for \textit{Swift}/XRT \citep{2007SPIE.6686E..07B}.
    The slow-moving objects also need to be taken into account especially when the field lies near the ecliptic plane. Some asteroids may be detected in VT images and can be verified using the Minor Planet Center (MPC\footnote{https://www.minorplanetcenter.net/}). 
    
    \item The brightness generally declines over time. While early-time behavior may exhibit complex features such as flares, plateaus, or rebrightenings, the overall trend of GRB optical afterglows at later phases (hours to days) is a decay in brightness.  This property is particularly useful in regions without DESI coverage, or where the depth of existing reference catalogs is insufficient.  Constructing light curves for all sources within the error box of the trigger or the X-ray counterpart is a practical method to identify the most promising candidates.

\end{itemize}

    There are some false positives in the VT images which might contaminate the candidates in the case of error box with an order of several arcminutes. The first effect arises from cosmic rays. 
    An uncataloged source detected in multiple single frames or in the stacked image is only considered to be a candidate of interest.
    The second effect is from hot pixels. The dark correction has been applied prior to the procedure of identification. However some new hot pixels may still emerge or the brightness of the existing hot pixels  may intensify, owing to the radiation damage from cosmic rays in space.  In the first year, internal calibration observations to acquire the dark calibration frames were performed with a frequency of once per month. During the gaps between consecutive calibration observations, we found some new hot pixels present in the calibrated images. These hot pixels could be removed by the image stacking in normal-stability mode, thanks to the drift between the images.
    Point spread function (PSF) and the hot pixels table are utilized for filtering,  particularly when a candidate is detected only in one channel.
 
    There is a wide distribution in the brightness of optical afterglows even at similar epochs after the bursts \citep{Kann2011}. 
   Bright afterglows can be identified in single frames. However, optical counterparts fainter than 22 magnitude can only be detected in stacked images. The differential imaging method is also used to detect the faint afterglow in cases that it overlaps with a bright host galaxy. For bright counterparts, the first and second halves of the initial orbit can be separated and stacked independently to generate a differential image. 
   For faint afterglows, templates for differential imaging are obtained by stacking images from the final orbits, which could introduce a long time gap between the image pairs used for differencing. However, in this latter case, the identification requires more time owing to the long delay needed for all science images to be downlinked via X-band passes. 
    
    Given the complexity of GRB optical light curves at the early phase, which can arise from reverse shocks, forward shocks, central engine activities, together with the wide distribution of optical afterglows, multiple methods must be employed synergistically to characterize the afterglows.

\subsection{Identification for bursts from ToO observations}

For bursts triggered by external missions or confirmed \textit{SVOM}/ECLAIRs triggers on the ground that do not automatically initiate a platform slew, ToO observations are scheduled. In most cases, the observation duration will be two orbits. Image stacking can achieve a depth of 23-24 mag. The configuration with the normal-stability mode of platform is suggested. The removal of hot pixels and cosmic rays is effectively accomplished in stacked images, owing to the shift during the observations in normal-stability mode. 
The typical downlink delay for ToO data is several hours. Better localization of X-ray counterparts reported by \textit{Swift}/XRT or  \textit{EP}/FXT greatly benefits the identification of candidate counterparts, if no optical counterpart is reported in GRB community\footnote{https://gcn.nasa.gov/circulars}.


\subsection{Identification for high-z GRB candidates}

One of the primary scientific objectives of VT is to select high-redshift GRB candidates through early, deep in-orbit optical observations covering the wavelength range up to 1 $\mu m$. As the Ly$\alpha$ line is redshifted to longer wavelengths, emission from sources at redshifts greater than 7 will be shifted beyond the wavelength coverage of VT’s red channel.

However, other scenarios may also produce optically dark bursts. One is intrinsically faint optical bursts with unusual microphysical parameters, for example those residing in a very low‑density external medium. The other is heavy line-of-sight absorption, as in the case of GRB 070306 \citep{Jaunsen2008}, GRB 080607 \citep{Perley2011}, GRB 090417B \citep{Holland2010}. 

Two conditions must be satisfied to ensure the effectiveness of this selection method: first, observations must begin within minutes of the gamma-ray burst and reach sufficient depth to avoid missing rapidly fading afterglows. Second, relatively accurate localizations of the X-ray counterparts are required, at the level of a few arcseconds, as achieved by \textit{Swift}/XRT or \textit{EP}/FXT. Since onboard images are easily affected by cosmic rays and other instrumental factors, the rapid and effective removal of these contaminants is crucial. Improving positioning accuracy can significantly reduce such artifacts, thereby enhancing the confidence level of the results. One successful application of this method is the identification of GRB 250314A \citep{Cordier2025}.

On the other hand, once an optical counterpart is detected in VT\_R or in both bands, its VT color can be used to estimate the redshift \citep{Wang2020}. The VT color redder than 2 can serve as an indicator for the redshift larger than 4, as performed and later confirmed by the spectroscopic observation in GRB 250215A\citep{2025GCN.39333....1X, 2025GCN.39343....1S}.

\section{Performance of identification and detection}

\begin{figure}
    \centering
    \includegraphics[width=1.0\linewidth]{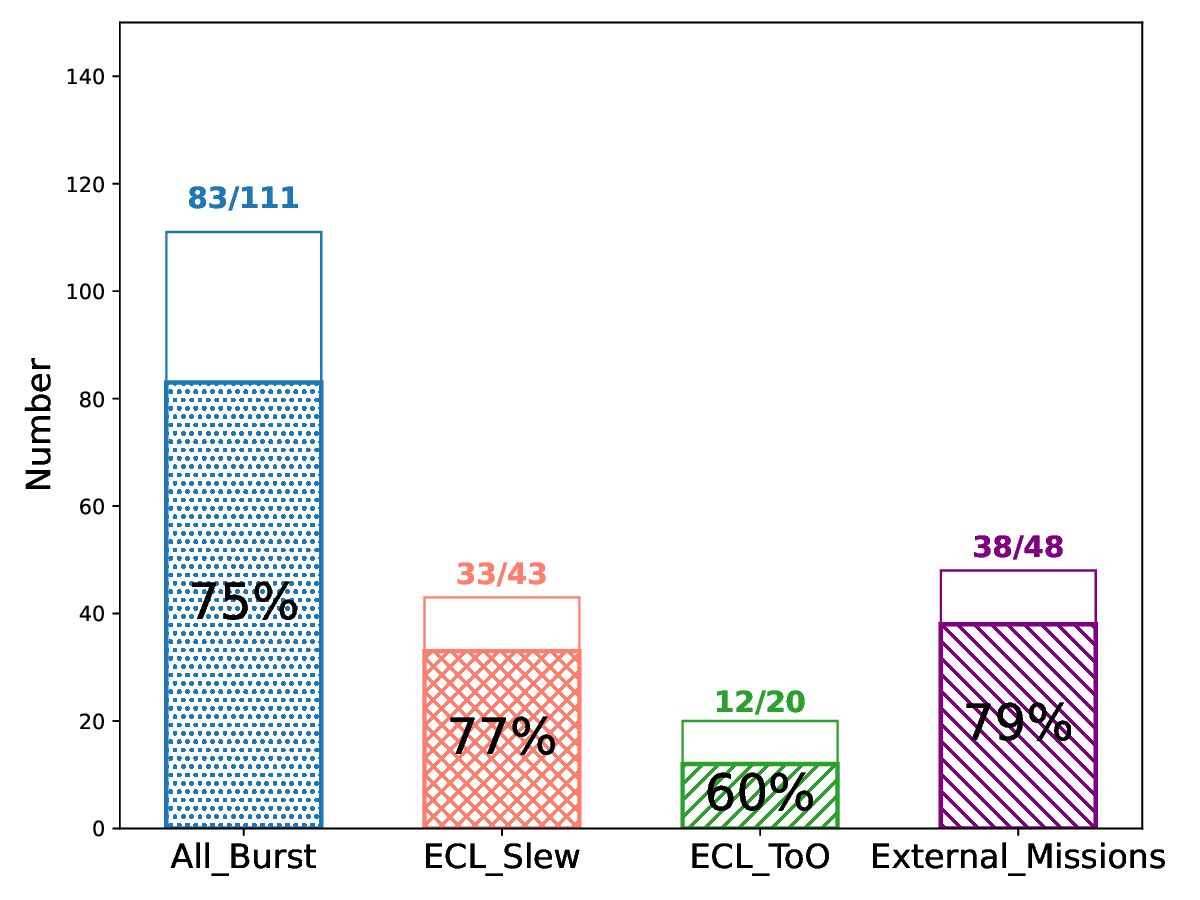}
    \caption{The detection rate in the VT X-band images for all the 111 GRBs observed by VT up to 2025 December 3, including ECLAIRs bursts with automatic slew (ECL\_Slew, see also Figure \ref{fig:ecl-slew}) and Table \ref{tab:ecl-slew}), ECLAIRs bursts with ToO observation (ECL\_ToO, see also Figure \ref{fig:ecl-ToO} and Table \ref{tab:ecl-nonslew}), and the bursts triggered by external missions (see also Figure \ref{fig:external-ToO} and Table \ref{tab:otherbursts}). }
    \label{fig:detectionrate}
\end{figure}

Up to 2025 December 3, 111 GRBs had been followed up by \textit{SVOM}/VT, in which 63 were triggered by \textit{SVOM}/ECLAIRs and 48 by external missions, including \textit{SVOM}/GRM \citep{Sun2026}, Einstein Probe \citep{Yuan2022}, the Neil Gehrels \textit{Swift} Observatory \citep{Gehrels2004}, Fermi/GBM \citep{Meegan2009}, INTEGRAL \citep{Winkler2003}, etc.. 
Overall, optical counterpart detection rate from VT X-band data is 75\%. The detailed detection rates for all GRBs that VT observed, as well as for each subgroup, are also shown in Figure \ref{fig:detectionrate} for comparison. 



\subsection{\textit{SVOM}/ECLAIRs bursts with automatic slew}

\begin{figure}
    \centering
    \includegraphics[width=1.0\linewidth]{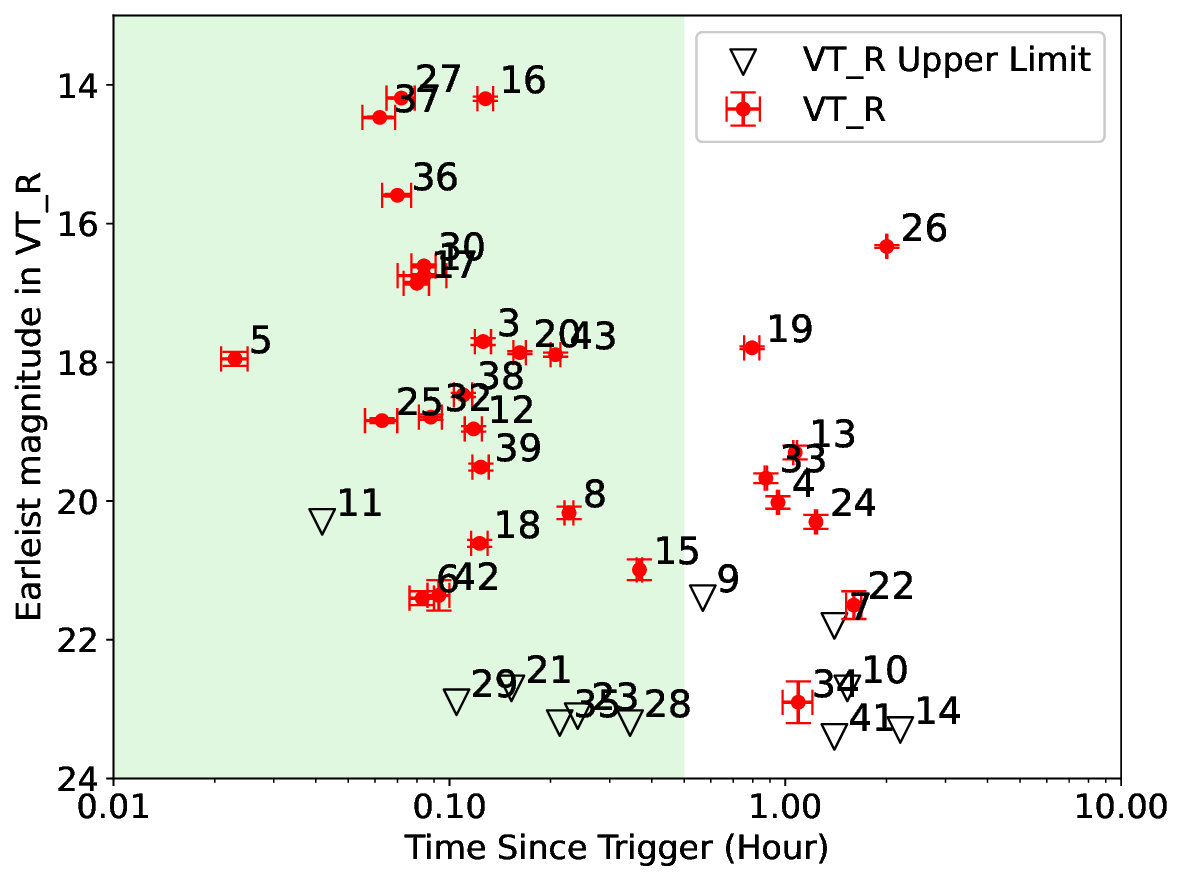}  
    \caption{The earliest VT\_R magnitudes of the optical counterparts for 43 \textit{SVOM}/ECLAIRs bursts with automatic slew onboard up to 2025 December 3. Numerical labels adjacent to the data points correspond to the IDs presented in Table \ref{tab:ecl-slew}. The shaded region on the left corresponds to observations with a mid-time of 30 minutes post trigger.}
    \label{fig:ecl-slew}
\end{figure}

By 2025 December 3, VT had observed 43 ECLAIRs-triggered bursts with automatic slew of the platform, as shown in Figure \ref{fig:ecl-slew} and Table \ref{tab:ecl-slew}. Here we only present the VT\_R brightness for clarity.  All detections are shown in red, while the upper limits are shown in black triangles. 

The brightnesses in VT\_R for ECLAIRs bursts with automatic slew range from 14.0 to $>$23.4 mag, with a median of VT\_R$\sim$18.9 mag. The mid-time of the observations ranges from 84 seconds to 2 hours, with a median value of 583.2 seconds (i.e., 9.72 minutes). For VT, the starting time (T$_{start}$) of images with good quality post trigger is influenced by the angle between the GRB and the platform pointing at that epoch, as well as by the time required for the platform to stabilize. Notably the start time could be delayed by more than one hour, mostly due to Earth occultation, since some bursts are triggered near the end of the best observing window in an orbit. 

Among the 43 ECLAIRs-triggered bursts with automatic slew that VT had observed, optical counterparts were detected for 33 GRBs, corresponding to a detection rate of about 77\%. Specifically,  27 ECLAIRs bursts were observed within the mid-time of 30 minutes post triggers, as shown in the shaded region on the left side of Figure \ref{fig:ecl-slew}. Of these, 21 GRBs had optical counterparts, resulting in a 77\% detection rate.

We also note that 10 bursts had no optical counterpart detected in VT X-band data, as shown in Figure \ref{fig:ecl-slew} with black inverted triangles. The VT\_R upper-limits of these bursts are listed in Table \ref{tab:ecl-slew}.
The possible explanations for these optical nondetections include intrinsic optical faintness in optical, heavy extinction,  or high-redshift GRBs. Early nondetections from VT X-band data could help to identify high-redshift GRB candidates. One successful application of this method is GRB 250314A (z=7.3). A detailed identification process through VT observation can be found in \citet{Cordier2025} and \citet{qiu2026}.

\subsection{\textit{SVOM}/ECLAIRs bursts with ToO observations}

\begin{figure}
    \centering
    \includegraphics[width=1.0\linewidth]{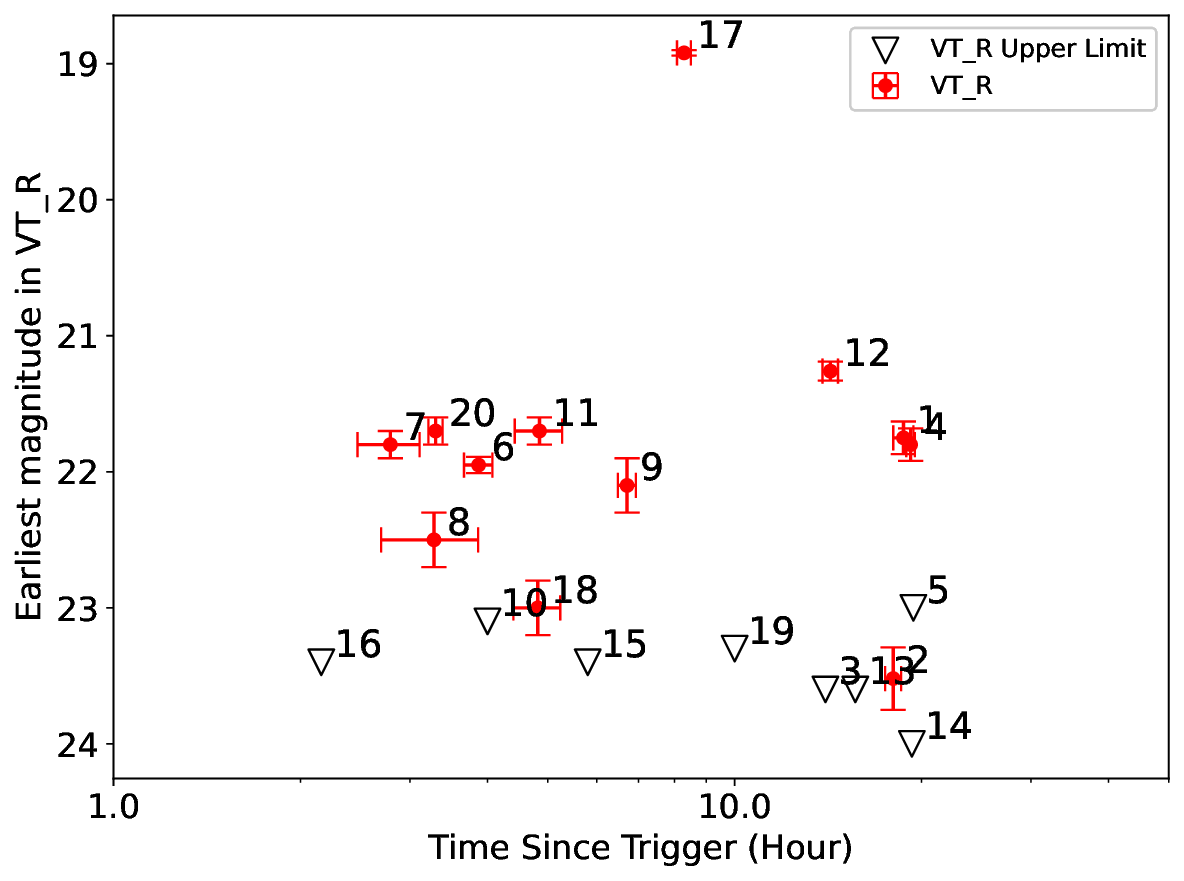}  
    \caption{The earliest VT\_R magnitudes of the optical counterparts for 20 \textit{SVOM}/ECLAIRs bursts with ToO observations up to 2025 December 3. Numerical labels adjacent to the data points correspond to the IDs presented in Table \ref{tab:ecl-nonslew}.}
    \label{fig:ecl-ToO}
\end{figure}

For ECLAIRs bursts that do not meet the onboard slewing criteria \citep{Schanne2026}, observations are performed in ToO mode. Up to 2025 December 3, 20 ECLAIRs bursts had been observed in ToO mode, as shown in Figure \ref{fig:ecl-ToO} and Table \ref{tab:ecl-nonslew}. Most of these bursts were detected from ECLAIRs onboard triggers, except for two events from offline ground-based triggers, which are marked with "$*$" in Table \ref{tab:ecl-nonslew}. All these 20 bursts had X-ray counterparts in \textit{SVOM}/MXT, \textit{Swift}/XRT, or \textit{EP}/FXT. 5 ECLAIRs ToO bursts were also detected by \textit{SVOM}/GRM instruments, which are marked with "$\lozenge$" in Table \ref{tab:ecl-nonslew}.

The observing times range from 2.16 hours to 19.20 hours, with a median of 5 hours post trigger. About 50\% of them were followed up by VT within 5 hours. The VT\_R magnitudes for the ECLAIRs ToO bursts range from 18.0 to $>$23.5 mag, with the median of VT\_R$\sim$18.9 mag. About 90\% of the counterparts of ECLAIRs ToO bursts are fainter than 21 mag in VT\_R. 

Of the 20 bursts, 12 had optical counterparts detected by VT, corresponding to a detection rate of only 60\%. The relatively low detection rate for these bursts via ToO observations can be attributed to at least two factors. First, the longer response delay via ToO observation, even though the Beidou Navigation Satellite System typically responds to requests within approximately 50 minutes. Second, most of these bursts have relatively low signal-to-noise ratios, which may be  associated with fainter optical counterparts.

\subsection{GRBs triggered by external missions}

\begin{figure}
    \centering
    \includegraphics[width=1.0\linewidth]{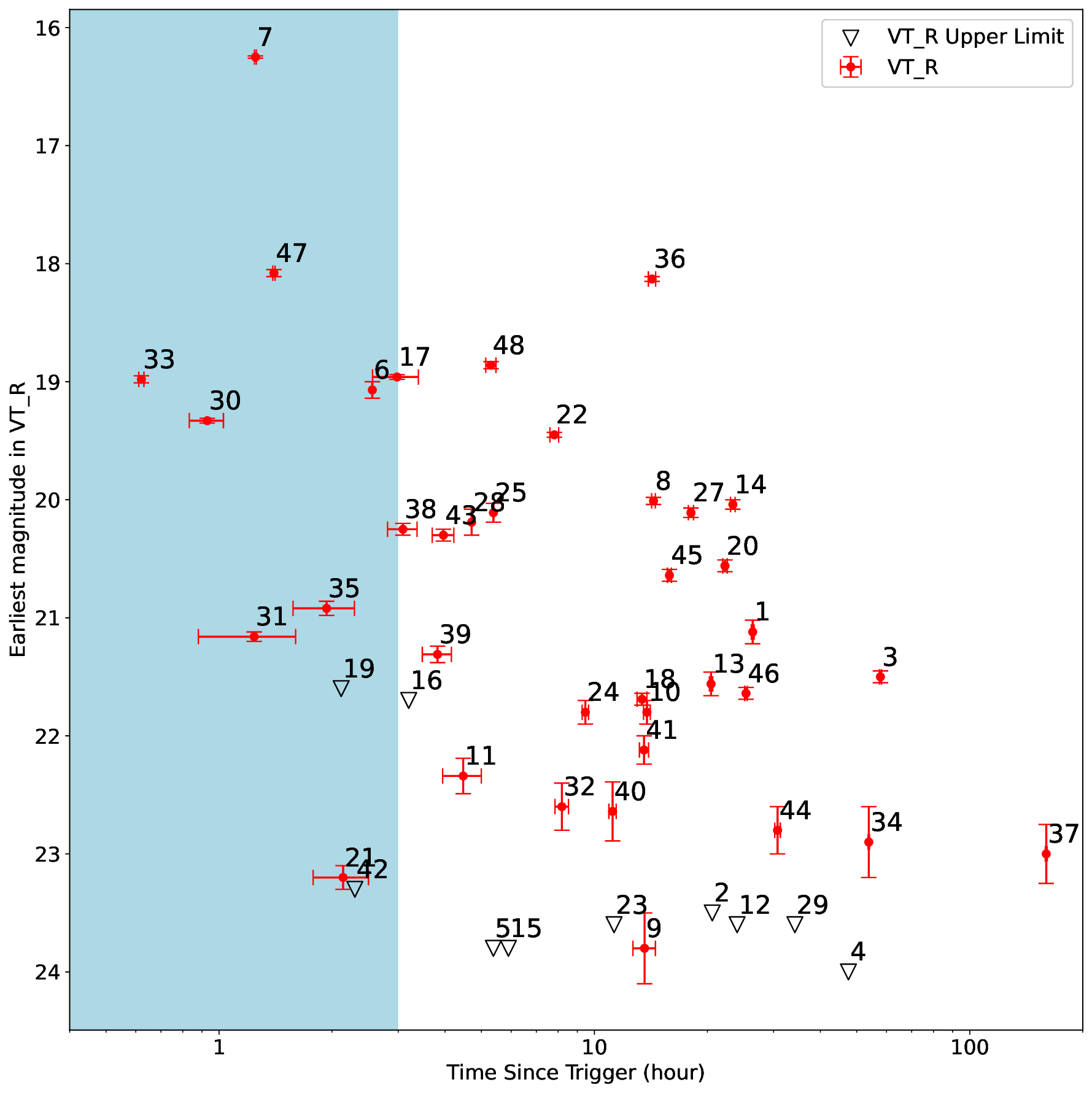}  
    \caption{The earliest VT\_R magnitudes of the optical counterparts for 48 bursts triggered by external missions with ToO observations up to 2025 December 3. Numerical labels adjacent to the data points correspond to the IDs presented in Table \ref{tab:otherbursts}. The shaded region on the left corresponds to observations with a mid-time of 3 hours post trigger.}
    \label{fig:external-ToO}
\end{figure}

For GRBs triggered by external missions that are followed up via VT ToO observations, two constraints must be satisfied: (1) Galactic latitude $>15^\circ$ (Configurable); (2) Sun avoidance angle $>90^\circ$. The typical start time for ToO observations for these bursts is approximately 50 minutes post trigger.
Up to 2025 December 3, VT has performed follow-up observation for 48 GRBs triggered by external missions, as illustrated in Figure \ref{fig:external-ToO} and listed in Table \ref{tab:otherbursts}.

The observing time ranges from 0.6 hours to 159.8 hours with a median of about 8.0 hours. The VT\_R magnitudes range from 16.25 to $>$23.8 mag, with a median of VT\_R$\sim$21.0 mag for these bursts.
The optical counterparts of 38 GRBs were detected by VT, corresponding to a detection rate of 79\%, which is comparable to the VT detection rate for \textit{SVOM}/ECLAIRs observed with automatic slew onboard and very short delays. This rate is at least 10\% higher than that for ECLAIRs triggers with ToO observations, which may be related to the different energy ranges of the respective instruments. This trend will be further investigated in the coming years with more observations.

Since May 2025, automatic ToO observation have performed through the Beidou system for triggers that meet the constraints mentioned above. No human intervention is required during this process \citep{Bai2026,Han2026}. If we consider only the GRBs with a mid-time of less than 3 hours, as shown in the shaded region on the left side of Figure \ref{fig:external-ToO}, 9 out of 11 have detected optical counterparts, giving a higher detection rate of 81\%. This improvement of detection is attributed to the rapid response of the Beidou uplink system.

\section{Summary and improvements in the future}

In this paper, we present the data processing of VT X-band data, mainly focusing on the GRB science, including the identification of GRB optical counterparts. Up to 2025 December 3, 
\textit{SVOM}/VT had performed observations for 111 GRBs mainly from \textit{SVOM}, \textit{Swift} and \textit{EP} missions. 
Among them, 63 bursts were triggered by ECLAIRs, including 43 ones with onboard automatic slew, and 48 GRBs were triggered by external missions. 

The overall VT detection rate of the optical counterparts of these GRBs is approximately 75\%. 
For SVOM/ECLAIRs bursts observed within 30 minutes post trigger, the detection rate is around 77\%. A slightly higher detection rate of 81\% is also achieved for bursts triggered by external missions with ToO observation within 3 hours of mid-time, albeit with a limited sample size.

Thanks to the robust data link from space to ground centers, the good performance of the data processing and the rapid response by operation teams.
The high detection rate of GRB optical counterparts, together with the efficient source identification,  demonstrates VT's contributions to the GRB community, especially for studies of the GRB population and optically faint bursts. To date, a series of studies have been conducted based on VT X-band data processing results, including observational analyses of GRB 240825A \citep{Wu2025}, GRB 241105A \citep{Dimple2025}, etc. Further details on SVOM GRB observations are provided in \citet{Daigne2026}.

In the future, further improvements will be implemented for the VT X-band data processing procedure, with a particular focus on targets located in complex backgrounds, such as those affected by contamination from nearby sources, local stray light, blooming, or pollution from detectors.

\begin{acknowledgements}

The Space-based multi-band astronomical Variable Objects Monitor (SVOM) is a joint Chinese-French mission led by the Chinese National Space Administration (CNSA), the French Space Agency (CNES), and the Chinese Academy of Sciences (CAS). We gratefully acknowledge the unwavering support of NSSC, IAMCAS, XIOPM, NAOC, IHEP, CNES, CEA, and CNRS.
This work is supported by the National Key R\&D Program of China (grant Nos. 2024YFA161170* and 2024YFA1611700) and by the National Natural Science Foundation of China (grant Nos. 12494571, 12494570, 12494573 and 12133003).
The authors are thankful for support from the Strategic Priority Research Program of the Chinese Academy of Sciences (Grant No.XDB0550401).


\end{acknowledgements}

\bibliography{2026-0041ref}
\bibliographystyle{aasjournal}


\appendix                  

\begin{table}[t]
   \caption{Earliest magnitude in VT\_R for the optical counterparts of 43 \textit{SVOM}/ECLAIRs bursts with automatic slew up to 2025 December 3 (see also Figure \ref{fig:ecl-slew}). For nondetections, 3$\sigma$ limit magnitude in VT$\_$R is given. For GRB 241030B,  '-' indicates that no measurement in VT\_R is available since the high background during the observations prevented the images from passing filtering criteria. But it was detected in VT\_B with the brightness of 19.47$\pm$0.05 mag at a mid-time of 382 seconds post trigger, as reported in GCN 37999.} 
   \centering
   \begin{tabular}{llrlll}
   \hline\hline
    ID & Burst  & Mid-Time  &  VT\_R   \\
    &   & (Hour)  &  (Mag)    \\
   \hline
   1 & GRB 241018A & 0.084 & 16.75$\pm$0.02       \\
   2 & GRB 241030B & 0.106  &  -                  \\ 
3 & GRB 241112B & 0.126 & 17.70$\pm$0.05        \\
4 & GRB 241212A & 0.951 & 20.02$\pm$0.09      \\
5 & GRB 241209B & 0.023 & 17.95$\pm$0.10       \\
6 & GRB 241217A & 0.083 & 21.40$\pm$0.10       \\
7 & GRB 241229A & 1.400 &  $>$21.8               \\
8 & GRB 250103A & 0.227 & 20.17$\pm$0.09         \\
9 & GRB 250106A & 0.568 & $>$21.4                \\
10 & GRB 250108A & 1.530 & 22.1$\pm$0.2         \\
11 & GRB 250127A & 0.042 & $>$20.3              \\
12 & GRB 250205A & 0.118 & 18.96$\pm$0.04       \\
13 & GRB 250215A & 1.070 & 19.30$\pm$0.10        \\
14 & GRB 250314A & 2.200 & $>$23.3                \\
15 & GRB 250317B & 0.368 & 20.99$\pm$0.15        \\
16 & GRB 250327B & 0.128 & 14.20$\pm$0.03       \\
17 & GRB 250329A & 0.080 & 16.86$\pm$0.02       \\
18 & GRB 250402A & 0.123 & 20.61$\pm$0.05      \\
19 & GRB 250419A & 0.796 & 17.79$\pm$0.02      \\
20 & GRB 250502A & 0.162 & 17.86$\pm$0.02      \\
21 & GRB 250507A & 0.153 & $>$22.7              \\
22 & GRB 250512B & 1.600 & 21.50$\pm$0.20    \\
23 & GRB 250530A & 0.241 & $>$23.1               \\
24 & GRB 250610B & 1.234 & 20.30$\pm$0.10      \\
25 & GRB 250612D & 0.063 & 18.84$\pm$0.03     \\
26 & GRB 250704A & 2.005 & 16.33$\pm$0.02       \\
27 & GRB 250706B & 0.072 & 14.19$\pm$0.01    \\
28 & GRB 250713A & 0.345 & $>$23.2              \\
29 & GRB 250806A & 0.105 & $>$22.9             \\
30 & GRB 250813B & 0.084 & 16.61$\pm$0.02       \\
31 & GRB 250818B & 0.076 & 17.3$\pm$0.1        \\
32 & GRB 250901A & 0.088 & 18.79$\pm$0.04      \\
33 & GRB 250903A & 0.875 & 19.67$\pm$0.07       \\
34 & GRB 250910A & 1.093 & 22.90$\pm$0.30       \\
35 & GRB 250912A & 0.213 & $>$23.2               \\
36 & GRB 251002A & 0.070 & 15.59$\pm$0.02       \\
37 & GRB 251013C & 0.062 & 14.47$\pm$0.01     \\
38 & GRB 251025B & 0.110 & 18.47$\pm$0.03   \\
39 & GRB 251026A & 0.124 & 19.51$\pm$0.05    \\
40 & GRB 251111A & 1.340 & 22.3$\pm$0.2      \\
41 & GRB 251116C & 1.400 & $>$23.4            \\
42 & GRB 251122A & 0.093 & 21.36$\pm$0.22      \\
43 & GRB 251129A & 0.207 & 17.89$\pm$0.03     \\   
   \hline  
    \hline
    \end{tabular}
   \label{tab:ecl-slew}
   \end{table}

\begin{table}[t]
   \caption{The earliest magnitudes in VT$\_$R for the optical counterparts of 20 \textit{SVOM}/ECLAIRs bursts with ToO observations up to 2025 December 3 (see also Figure \ref{fig:ecl-ToO}). "$\lozenge$" refers to the detection by both \textit{SVOM}/ECLAIRs and \textit{SVOM}/GRM. "$*$" refers to the bursts from \textit{SVOM}/ECLAIRs offline ground-based triggers.
   }
   \centering
   \begin{tabular}{llrlll}
   \hline\hline
    ID & Burst  & Mid-Time  &  VT\_R   \\
       &   & (Hour)  &  (Mag)    \\
   \hline
1 & GRB 240821A  $\lozenge$   & 18.700 &  21.75$\pm$0.12   \\
2 & GRB 241001A               & 18.000 &  23.52$\pm$0.23   \\
3 & GRB 241017A               & 14.000 &  $>$23.6          \\  
4 & GRB 241029A $\lozenge$    & 19.200 &  21.80$\pm$0.12   \\
5 & GRB 241104A $\lozenge$$*$ & 19.400 &  $>$23.0          \\
6 & GRB 241113B $\lozenge$    & 3.870  &  21.95$\pm$0.06   \\
7 & GRB 250103B               & 2.790  &  21.80$\pm$0.10   \\
8 & GRB 250213A               & 3.280 &  22.50$\pm$0.20    \\
9 & GRB 250219A               & 6.710 &  22.10$\pm$0.20    \\
10 & GRB 250329B              & 4.000 &  $>$23.0           \\
11 & GRB 250328A              & 4.850 &  21.70$\pm$0.10    \\
12 & GRB 250403A              & 14.260 &  21.26$\pm$0.07   \\
13 & GRB 250506A $\lozenge$   & 15.640 &   $>$23.6          \\ 
14 & GRB 250521B $*$          & 19.290 &  $>$24.0          \\
15 & GRB 250628B              & 5.800 &  $>$23.4           \\
16 & GRB 250727A              & 2.160 &  $>$23.4           \\
17 & GRB 250812A              & 8.290 &  18.92$\pm$0.02    \\
18 & GRB 251007B              & 4.820 &  23.00$\pm$0.20    \\
19 & GRB 251116B              & 10.000 &  $>$23.3          \\
20 & GRB 251203C              & 3.300 &  21.70$\pm$0.10    \\
   \hline  
    \hline
    \end{tabular}
   \label{tab:ecl-nonslew}
   \end{table}

\begin{table}[t]
   \caption{The earliest magnitudes in VT$\_$R for the optical counterparts for 48 GRBs triggered by external missions up to 2025 December 3 (see also Figure \ref{fig:external-ToO}). For GRB 250625A, "-" indicates that no measurement is available in VT\_R since this target was contaminated by the blooming of bright stars nearby. But it was detected in VT\_B with the brightness of 22.50$\pm$0.15 mag at a mid-time of 1.15 hours  post trigger, as reported in GCN 40830.
   %
   }
   \centering
   \begin{tabular}{llrlll}
   \hline\hline
  ID & Burst  & Mid-Time  &  VT\_R   \\
      &   & (Hour)  &  (Mag)    \\
   \hline
  1 & GRB 240825A         &       26.400 &       21.12$\pm$0.10       \\
  2 & GRB 240905E         &       20.600 &       $>$ 23.5             \\
  3 & GRB 240910A         &       57.800 &       21.50$\pm$0.05       \\
  4 & GRB 240919A/EP240919a  &    47.500 &       $>$ 24.0             \\
  5 & GRB 241006A         &       5.380  &       $>$ 23.8             \\
  6 & GRB 241025A         &       2.560  &       19.07$\pm$0.07       \\
  7 & GRB 241030A         &       1.250  &       16.25$\pm$0.01       \\
  8 & GRB 241105A         &       14.366 &       20.01$\pm$0.03       \\
  9 & GRB 241213A        &        13.600 &       23.80$\pm$0.30       \\
 10 & GRB 241229B        &        13.800 &       21.80$\pm$0.10       \\
 11 & GRB 250109A/EP250109a  &    4.470  &       22.34$\pm$0.15       \\
 12 & GRB 250114B        &        24.000 &       $>$ 23.6             \\
 13 & GRB 250226A/EP250226a  &    20.450 &       21.56$\pm$0.10       \\
 14 & GRB 250309B/ZTF25aaitvjt  & 23.376 &       20.04$\pm$0.04       \\
 15 & GRB 250327A        &        5.900  &       $>$ 23.8             \\
 16 & GRB 250330A        &        3.200  &       $>$ 21.7             \\
 17 & GRB 250424A        &        2.980  &       18.96$\pm$0.02       \\
 18 & GRB 250430A        &        13.390  &       21.69$\pm$0.05      \\
 19 & GRB 250504A        &        2.780  &         $>$23.0             \\
 20 & GRB 250515A/GOTO25cqo  &   22.270   &       20.56$\pm$0.05      \\
 21 & GRB 250520A        &        2.140  &        23.2 $\pm$0.1        \\
 22 & GRB 250521D        &        7.819  &       19.45$\pm$0.02       \\
 23 & GRB 250605A        &       11.280  &       $>$ 23.6             \\
 24 & GRB 250612B/EP250612a     &        9.460  &   21.80$\pm$0.10    \\
 25 & GRB 250617B        &        5.380  &       20.11$\pm$0.08       \\
 26 & GRB 250625A        &        1.150  &         -                   \\
 27 & GRB 250702F        &        18.080  &       20.11$\pm$0.04      \\
 28 & GRB 250704B/EP250704a      & 4.709  &       20.19$\pm$0.11      \\
 29 & GRB 250702B/EP250702a      & 34.200 &        $>$ 23.6           \\
 30 & GRB 250725A        &        0.930  &       19.33$\pm$0.02       \\
 31 & GRB 250805A        &        1.240  &       21.16$\pm$0.04       \\
 32 & EP250806a          &        8.190  &       22.60$\pm$0.20       \\
 33 & GRB 250807A        &        0.621   &       18.98$\pm$0.03      \\
 34 & GRB 250807B        &        53.830  &       22.90$\pm$0.30      \\
 35 & GRB 250823A        &       1.934    &       20.92$\pm$0.06      \\
 36 & GRB 250919A/EP250919a   &  14.240   &       18.13$\pm$0.02      \\
 37 & GRB 250916A        &      159.840   &       23.00$\pm$0.25      \\
 38 & GRB 250924A        &        3.089   &       20.25$\pm$0.05      \\
 39 & GRB 251001B        &        3.819   &       21.31$\pm$0.07      \\
 40 & GRB 251005C        &        11.180  &       22.64$\pm$0.25      \\
 41 & GRB 251006B        &        13.560  &       22.12$\pm$0.12      \\
 42 & GRB 251011B        &        2.300   &       $>$ 23.3            \\
 43 & GRB 251017A        &        3.960   &       20.30$\pm$0.05      \\
 44 & GRB 251021A/EP251021b    &  30.750  &       22.80$\pm$0.20      \\
 45 & GRB 251118C/EP251118a    &  15.840  &       20.64$\pm$0.05      \\
 46 & GRB 251126A        &        25.340  &       21.64$\pm$0.05      \\
 47 & GRB 251201B        &        1.400   &       18.08$\pm$0.03      \\
 48 & GRB 251202A/EP251202a   &   5.300   &      18.86$\pm$0.03       \\
   \hline  
    \hline
    \end{tabular}
   \label{tab:otherbursts}
   \end{table}

\label{lastpage}

\end{document}